\documentclass[traditabstract]{aa} 
\usepackage{graphicx}
\usepackage{txfonts}

\usepackage{natbib} 
\bibpunct{(}{)}{;}{a}{}{,} 
\usepackage{url}

\newcommand{\source}{\mbox{1H0419-577}}

\def \th {\thinspace}
\def \kms {\hbox{km s$^{-1}$}}

\def \ergsc{\hbox{erg s$^{-1}$ cm$^2$}}

\def \ergs{\hbox{erg s$^{-1}$}}

\def \msun {\hbox{${\rm M_\odot}$}}
\def \msunyr{\hbox{{$M_{\odot}$ yr$^{-1}$}}}

\newcommand\omegam{\hbox{{$\Omega_{\rm m}$}}}
\newcommand\omegalambda{\hbox{{$\Omega_{\Lambda}$}}}
\newcommand\kmsmpc{{\rm km s$^{-1}$ Mpc$^{-1}$}}
\newcommand\ho{\hbox{{$H_{0}$}}}
\def \nh {\hbox{ $N{\rm _H}$ }}
\def \colc {cm$^{-2}$}

\def \fwhm {{\em FWHM}}
\newcommand{ \lia} {Ly$ \rm{\,\sc{\alpha}}$}
\newcommand{\xmm }{{\rm XMM-\it{Newton}}}
\newcommand{\epic }{{\rm EPIC}}
\newcommand{\rgs }{{\rm RGS}}

\newcommand{\hst }{{\rm HST}}
\newcommand{\Cos }{\rm HST-COS}

\newcommand{\xabsinput }{{\it xabsinput}}
\newcommand{\rgsfluxcombine }{{\it RGS\_fluxcombine }}
\newcommand{\rgsfmat }{{\it RGS\_fmat }}

\newcommand{\rgsfluxer }{{\sf rgsfluxer }}
\def \chirid{$\chi ^{2}_{\nu}$}

\begin{document}

  \title{Simultaneous XMM-\textit{Newton} and HST-COS observation of 1H0419-577:}

 \subtitle{the absorbing and emitting ionized gas.}

  \author{L. Di Gesu \inst{1}
      \and E. Costantini \inst{1}
      \and N. Arav\inst{2}
       \and B. Borguet \inst{2}     
      \and R.G. Detmers \inst{1}
      \and J. Ebrero\inst{1}
     \and D. Edmonds\inst{2}
     \and J.S. Kaastra\inst{1}
      \and E. Piconcelli\inst{3,4}
    \and F. Verbunt\inst{1,5}
}

\institute{
SRON Netherlands Institute for Space Research, Sorbonnelaan 2, 3584 CA Utrecht, The Netherlands \email{L.di.Gesu@sron.nl}
\and Department of Physics, Virginia Tech, Blacksburg, VA 24061, USA
\and Osservatorio Astronomico di Roma (INAF), Via Frascati 33, I--00040, Monteporzio Catone (Roma), Italy
\and XMM-Newton Science Operations Centre, ESA, PO Box 78, 28691 Villanueva de la Canada, Madrid, Spain
\and Department of Astrophysics/IMAPP, Radboud University, PO Box 9010, 6500 GL Nijmegen, The Netherlands
}


\abstract
{
In this paper we analyze the X-ray, UV and optical data of the Seyfert 1.5
galaxy 1H0419-577, with the aim of  detecting and studying an
ionized-gas outflow. The source was observed  simultaneously 
in the X-rays with \th  \xmm \th and in the UV with \th \Cos.
Optical data were also acquired with the \th \xmm \th Optical
Monitor. We detected a thin, lowly ionized warm absorber
($ \log \xi \approx 0.03$,
$\log \nh \approx 19.9$ \colc)
in the X-ray spectrum,
consistent to be produced by the same outflow already detected in the UV.
Provided the gas density estimated in the UV,
the outflow is consistent to be located in the host galaxy, at $\approx$ kpc scale.
Narrow emission lines were detected in the X-rays, in the UV and also in the
optical spectrum. A single photoionized-gas model cannot account for all the
narrow lines emission, indicating
that the narrow line region is probably a stratified environment,
differing in density and ionization.
X-ray lines are unambiguously produced
in a more highly ionized gas phase than the one emitting the UV lines.
The analysis suggests also that the X-ray emitter may be
just a deeper portion of the same gas layer producing
the UV lines. Optical lines are probably produced
in another, disconnected gas system. The different ionization
condition, and the $\approx$ pc scale location 
suggested by the line width for the narrow lines emitters,
argue against a connection between the warm absorber and the narrow line
region in this source.
}
\titlerunning{Simultaneous XMM-\textit{Newton} and HST-COS observation of 1H0419-577}
\authorrunning{L. Di Gesu et al}
 \maketitle
%

%
%
\section{Introduction} 
\label{Introduction} 
Active galactic nuclei (AGN)
are believed to be powered
by the accretion of matter
onto a supermassive black hole
\citep{ant1993}. Emission
and absorption lines in AGN
spectra are the signatures
of a plasma photoionized
by the central source. 
High-resolution spectral 
observations can
probe the physical conditions 
in this gas, providing 
information about the interaction 
between AGN radiation and the
surrounding environment.
\\  \\
%
%
%
A variety of emission lines
with different width
(from $10^{2}$ to $10^{4}$ \kms)
can be identified in AGN 
type 1 spectra.
The diverse
line-broadenings reflect
a different location of the emitting region:
narrower lines are believed to originate
in a region farther from
the black hole ($\approx$100 pc), 
lower in density and
in temperature 
than the broad-line 
region (BLR) where
broader lines are emitted
\citep{ost1989}.
Lines emission
ranges
from the optical
(e.g. 
[\ion{O}{iii}],
hydrogen Balmer series) 
to the X-ray waveband 
(e.g
\ion{O}{vii},
\ion{O}{viii},
\ion{Ne}{ix}).
Narrow lines may be more difficult
to detect.
For instance,
in the UV, narrow lines (NL)
from e.g. \ion{C}{iv},
\ion{O}{vi} are blended
in a dominant broad
component and difficult
to disentangle
\citep[e.g.][]{kri2011}. 
In the soft X-ray the
flux of any emission line
is usually outshone
by the underlying continuum
and lines detection
is favored
by a temporary low-flux state
of the source \citep[e.g. NGC 4051,][]{nuc2010}.
Whether X-ray lines
arise in the same gas
emitting the longer-wavelength
lines is an open issue
that has been
recently addressed
through multiwavelength
photoionization modeling.
In the case of Mrk 279,
X-ray broad lines
are consistent to be produced
in a highly-ionized skin
of the UV and optical
BLR \citep{cos2007}.
\citet{bia2006}
found that a single
medium, photoionized
by the central continuum, 
may produce the
[\ion{O}{iii}] to soft X-ray
ratio observed in spatially
resolved images of the
narrow line region (NLR).
\\  \\
Besides the lines-emitting
plasma, another photoionized-gas component
that can modify
the spectra of type 1
AGN is a warm absorber
(WA), intervening in the
line of sight.
WA are commonly detected
in about half type 1 AGN 
\citep{cre1997,pic2005}
via UV and/or X-ray 
absorption lines.
These lines are
usually blueshifted,
\citep[see][]{cre2003}
indicating a global
outflow
of the absorber.
In the last ten years, 
multiwavelength UV-X-ray
campaigns (see
\citet{cos2010} and references therein)
have depicted
the physical conditions in the
outflowing gas with great
detail. WA
are multi-component
winds spanning a wide range in ionization
and in velocity \citep[e.g.][]{kri2011,ebr2011}.
X-ray spectra show the the most highly ionized lines
(e.g from \ion{O}{vii}, \ion{O}{viii} and
\ion{Ne}{ix}), 
while a lower-ionization phase (e.g. \ion{C}{ii}, \ion{Mg}{ii}) 
is visible only in the UV. 
In some cases (e.g. NGC 3783, NGC 5548,
NGC 4151, Mrk 279), a common phase
producing e.g. \ion{O}{vi} lines both in
the UV and in the X-ray spectrum has been
identified
\citep{gab2002,ste2005,kra2005,ara2007}.
Similarities in the line width 
\citep[NGC 3783,][]{beh2003}
and in the gas column density
\citep[NGC 5548,][] {det2010}
suggests  a connection between
the WA and the gas in the NLR.
However, the origin of WA
is not clearly established yet, 
mainly because of the great
uncertainty in estimating
its location.
\\ \\
Provided the gas density,
the distance
of the outflow from the
central source can be in principle
derived from the gas 
ionization parameter
$\xi=L_{\rm ion}/nR^2$
(where $L_{\rm ion}$ 
is the source ionizing luminosity,
$n$ is the gas density,
and $R$ is the distance 
from the ionizing source).
However, in most 
cases the gas density
is unknown and the distance may just be
estimated indirectly \citep{blu2005}.
Absorption lines from collisionally-excited
metastable levels
may provide a direct density 
diagnostic \citep[e.g.][]{bau2009},
but they are rarely detected.
In the UV, metastable lines from e.g. \ion{Fe}{ii}
\ion{Si}{ii} and \ion{C}{ii} have been
detected in a handful of cases
\citep[e.g.][]{moe2009,dun2010a,bor2012}. 
In the X-rays the identification
of metastable lines from \ion{O}{v}
is hampered by uncertainties
in the predicted line wavelength
\citep{kaa2004}.
So far,
available estimations,
using different methods,
have located the
outflows within the BLR \citep[e.g. NGC 7469,][]{sco2005}
or as far as the putative torus
\citep[e.g.][]{ebr2010}. 
In some quasars  a galactic-scale
distance have been reported.
\citep{ham2001,hut2004,bor2012,moe2009}.
\\ \\
The distance estimation allows
to quantify the amount
of mass and energy released
by the outflow into the medium.
Hence, it is possible
to establish if WA
contribute in the so-called
AGN feedback, which
is often invoked to
explain the energetics and
and the chemistry of the
medium up to a very large scale
\citep{sij2007,hop2008,som2008,mcn2012}.
Warm absorbers 
usually produce a negligible
feedback \citep[e.g.][]{ebr2011}.
Only the fastest AGN wind
\citep[e.g.][]{moe2009,dun2010a,tom2012},
may be dynamically
important in the evolution
of the interstellar medium
\citep{fau2012}.
\\ \\
In this
paper we analyze a long-exposure
\th \xmm \th  dataset of the
bright Seyfert galaxy \th \source \th, taken
simultaneously with
a Cosmic Origins Spectrograph
(\th \Cos \th) spectrum.
The source is a radio
quiet quasar located at redshift
z=0.104 and spectrally
classified as a type 1.5 Seyfert
galaxy \citep{ver2006}. Using the H$\beta$
line width, \citet{pou2004a} derived
a $1.3 \times 10^{8}$ \msun \th
mass for the supermassive black hole (SMBH)
hosted in the nucleus.
The HST-COS
spectrum has been published by 
\citet[][herafter E11]{edm2011}. 
The UV spectrum
displays broad emission lines
from \ion{C}{iv},  \ion{O}{vi}, 
and \th \lia \th as well as absorption 
lines (E11). Three outflowing components
were identified in absorption, 
with the line centroids
located at
$\rm v_1=-38 \,  \kms$, 
$\rm  v_2=-156 \,\kms$
and $\rm  v_3=-220 \, \kms$
in the rest frame of the source.
Only a few ionized species
(\ion{H}{i}, \ion{C}{iv}, \ion{N}{v},
\ion{O}{vi}) were detected in 
component 1, while component
2+3 displays a handful of transitions
from both low (e.g. \ion{C}{ii})
and high-ionization
species (e.g. \ion{C}{iv}, \ion{N}{v}, \ion{O}{vi}).
\\ \\
The present analysis is focused
mainly on the high-resolution
spectrum collected with the 
Reflection Grating Spectrometer \citep[RGS,][]{den2001};
in a companion paper we will present
the broad band X-ray spectrum of this source.
Analysis of a previous RGS dataset
is reported in
\citet{pou2004b}.
Hints of narrow absorption features
from an Fe unresolved transition array
(UTA) were noticed in the 
spectrum. However, the short exposure time 
($\approx 15$ ks)
prevented so far
an unambiguous detection and
characterization
of a  warm absorber
in this source.
\\ \\
The paper is organized as follows:
in Sect. \ref{obs} we present the
\th \xmm \th  observations and the data reduction;
in Sect. \ref{sed} we describe the spectral
energy distribution of the source; in Sect.
\ref{rgs} we discuss the spectral
analysis; in Sect. \ref{cfr_em} 
we model the narrow
emission lines of the source; 
in Sect. \ref{cfr_abs} we compare 
the X-ray and the UV absorber:
finally in Sect. \ref{disc} we discuss our results
and in Sect. \ref{sum} we present the conclusions.
The cosmological parameter used are:
\ho=70 \kmsmpc, \omegam=0.3, \omegalambda=0.7.
Errors are quoted at 68\% confidence levels
($\Delta \chi^2=1.0$) unless otherwise stated.

%
\section{Observations and data preparation}
\label{obs}
%
%
\begin{table}
\caption{XMM-Newton observation log for \th \source \th}     
\label{obs.tab}      
\centering                    
\begin{tabular}{l c c}        
\hline\hline                 
Observation ID  & 0604720401 & 0604720301\\
\hline 
Date & 28/05/2010 & 30/05/2010  \\
Orbit & 1917 & 1918 \\
Net exposure (ks) \tablefootmark{a} & 61 & 97  \\
\th \rgs \th Count Rate ($\rm s^{-1}$) & $0.337\pm 0.002$ & $0.385\pm 0.002$  \\ 
\th \epic-pn Count Rate ($\rm s^{-1}$)& $1.41 \pm 0.02$ & $1.56 \pm 0.02 $ \\              
  \hline                             
\end{tabular}
\tablefoot{
\tablefoottext{a}{Resulting exposure time after correction for flaring.}
}
\end{table}
%
%

In May 2010, two consecutive exposures 
of \th \source \th were taken with the
\th \xmm \th X-ray telescope both with
the EPIC
cameras \citep{stru2001,tur2001} and the RGS.
Moreover Optical Monitor \citep[OM,][]{mas2001} Imaging
Mode data were acquired with four broad-band
filters (B, UVW1, UVM2, UWV2) and
two grism filters (Grism1-UV and
Grism2-visual). The source was observed
for $\approx$ 167 ks in total and a slight
shift in the dispersion direction was applied
in the second observation. In this way,
the bad pixels of the RGS
detectors were not at the same location
in the two observations,
allowing us to correct them
by a combination of the two spectra
(see Sect. \ref{rgsd}).
Details of the two observations are provided
in Table \ref{obs.tab}.

\subsection{The RGS data}
\label{rgsd}

For both RGS datasets, we processed the data following 
the standard procedure
\footnote{\url{http://xmm.esac.esa.int/sas/8.0.0/documentation/threads/}},
using the \th \xmm \th Science
Analysis System (SAS, version 10.0.0) and the latest
calibration files. We created calibrated event
files for both RGS1 and RGS2, and
to check the variation of the background, 
we created also the  background light curves
from CCD 9. The background of 
the first observation
was quiescent, while the second light curve
showed contamination by soft protons flares. 
We cleaned the contaminated observation 
applying a time filter to the event-files: for this purpose,
we created the good time intervals (GTI) cutting in the light curve 
all the time bins where the count-rate was over 
the standard threshold of 0.2 counts s$^{-1}$. Resulting 
exposure time after deflaring is 97 ks.
Starting from the cleaned event files, we created a fluxed spectrum for
each RGS detector for both observations. We did
this through the SAS task \textsf{rgsfluxer}, considering
only the first spectral order and taking the background 
into account.\\
The spectral fitting for \th \source \th
was performed using the 
package SPEX, version 2.03.02 \citep{kaa1996}.
We first fit the RGS spectrum
of each observation to check whether the continuum
was unchanged in the two datasets: we found
that in both datasets the continuum
could be phenomenologically fitted by a broken
power law. The variations of the fitted parameters
were within the statistical errors. Moreover
the measured flux in the RGS bandpass was
basically the same (within $\approx$ 8\%) 
in the two observations.
This source is well known in the literature
for being highly variable in the
soft X-ray band ($<2.0\, \rm keV$)
\citep[e.g.][]{pou2004a}: our observation
caught it in an intermediate flux state
($F^{\rm pn}_{0.5-2.0 \, \rm keV} \approx 10^{-11}\,$ \ergsc).\\
Given the stability of the continuum shape
in the two observations, we could safely
perform a combination of the spectra,
to improve the signal to noise ratio.
We followed the same route outlined in
\citet{kaa2011};
here we just summarize the main steps.
We combined the four fluxed spectra
created with \th \rgsfluxer \th 
into a single stacked spectrum
using the SPEX auxiliary
program \th \rgsfluxcombine.  
The routine\th \rgsfluxcombine \th sums up 
two spectra using the exposure time to weigh all the
bins without bad pixels. 
In the presence of a bad pixel, 
this procedure is incorrect 
since it would create an artificial 
absorption line
\citep[for an analytical example see][]{kaa2011}.
To correct the output spectrum from
bad pixels, the task works as follows: 
in the presence of a bad pixel,
it looks at the neighbouring pixels and, 
assuming that the spectral shape
does not change locally, it estimates the
contribution to the total flux expected
from good data. Using this fraction,
the task estimates the factor 
by which the
flux at the bad pixel location
has to be scaled.
For the final composite spectrum,
the proper SPEX readable response 
matrix was hence created
through the tool \th \rgsfmat.
\subsection{The OM data}
We retrieved the processing pipeline
subsystem (PPS) products of the
OM Image Mode operations to
extract the source count-rates
in four broad band filters:
B ($\rm\lambda_{eff}$=4340 \AA),
UVW1 ($\rm\lambda_{eff}$=2910 \AA),
UWM2 ($\rm \lambda_{eff}$=2310 \AA),
and UVW2 ($\rm \lambda_{eff}$=2120 \AA).
Hence, we converted the count-rates
to flux densities 
using the conversion factors provided
in the SAS watchout web-page
\footnote{\url{http://xmm.esac.esa.int/sas/11.0.0/watchout/Evergreen_tips_and_tricks/uvflux.shtml}}.\\
We also processed the images from the OM optical
grism using the standard reduction pipeline 
(\textsf{omgchain})
provided in the SAS.
The task corrects the raw OM grism
files from the Modulo-8 fixed
pattern noise and removes
the residual scattered light features.
It rotates the
images aligning the grism dispersion axis
to the pixel readout columns of the images 
and it runs a source detection algorithm.
Finally, the tool extracts the spectra of
the detected sources
from the usable spectral orders.
A step by step description 
of the grism extraction chain 
is given in the SAS User's Guide.
For the present analysis we used the 5 ks long
optical spectrum 
(grism1, first dispersion order) from
the dataset 0604720301 to measure
the luminosities of the optical
narrow emission lines of \th
\source \th (Sect. \ref{optem}).
\subsection{The EPIC data}
To constrain the spectral energy distribution (SED)
of the source in the X-ray band we used the broadband EPIC-pn
spectrum. We applied the standard SAS data analysis thread 
to the observation data files (ODF)
products of both observations. We
created the calibrated EPIC-pn event files  
and we cleaned them from soft proton flares
through a time-filtering of the light curves.
The counts threshold over which we discarded
the time bins of the light curves
was determined by a 2$\sigma$
clipping of the light curve
in the whole EPIC band. 
Starting from
the clean event files we extracted
the source and background spectra and we created
the spectral response matrices.
We fitted the EPIC-pn spectrum
with a phenomenological model,
taking the galactic absorption
into account. The unabsorbed 
phenomenological continuum model
of the EPIC-pn spectrum
served as input for the X-ray SED.
%

\section{The spectral energy distribution}
\label{sed}
\begin{figure}[t]
\includegraphics[width=0.5\textwidth]{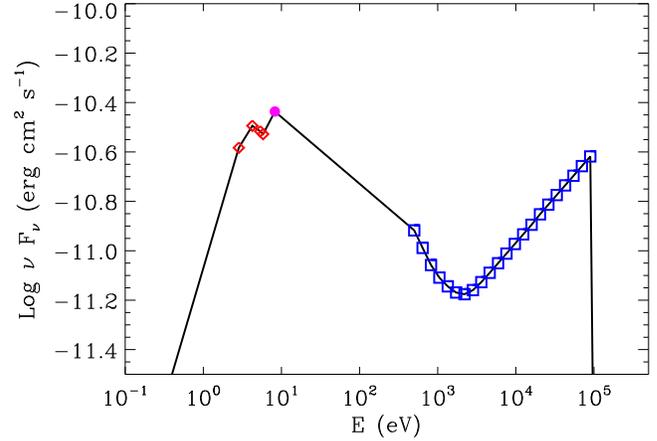}
\caption{The spectral energy distribution for \th \source \th.
Open diamonds are the fluxes from the OM optical and UV
broad band filters discussed in the text; the filled circle is the
continuum flux taken from the simultaneous \th \Cos \th
observation; open squares are from the unabsorbed model
for the X-ray continuum, obtained from the EPIC-pn
observation.}
\label{sed.fig}
\end{figure}
%
%
We exploited the simultaneous \th \xmm \th
and \th \Cos \th observation to constrain
the SED of
\th source. The shape of the SED at lower
energy is constrained by the \th \xmm \th
OM and by \th \hst \th, while the
\th \epic-pn camera covers the higher energy range.
For this purpose we estimated the UV continuum
of the source at 1500 \AA\
in the complete \th \Cos \th
spectrum.
We selected a wavelength region
not contaminated by any emission lines 
(1495 \AA--1505 \AA) and
we took the average value of the 83
spectral points comprised in it.\\
We cut off the SED at low ($\rm E \leq 1.36 \, eV$) 
and high energy ($\rm E \geq 100 \, keV$).
Indeed, the AGN spectral energy distribution falls
off with the square of the energy above
100 keV while the optical-UV bump has an
exponential cutoff in the infrared
\citep{fer2003}.
We show the SED
in Fig. \ref{sed.fig}.
From a numerical integration of
the SED we calculated the
source bolometric luminosity
($\rm \log L_{\rm bol}=45.95$ \ergs):
for the SMBH hosted in \th \source \th
the Eddington ratio is therefore
$L_{\rm bol}/ L_{\rm Edd}=0.5$.
The source ionizing luminosity
in the 1-1000 Ry energy range
is $\rm \log L_{\rm ion}=45.43$ \ergs.
%
%
\section{Spectral analysis}
\label{rgs}
 \begin{figure}[t]
 \includegraphics[angle=-90,width=0.5\textwidth]{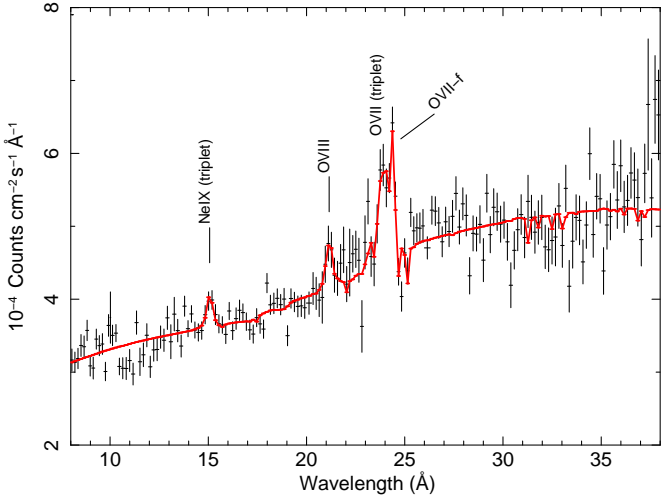}
    \caption{Composite RGS spectrum of \th \source,
             rebinned for clarity.
             The solid line corresponds to the best fit model.
             The main emission features are labeled.
            }
       \label{rgs_sp.fig}
 \end{figure}

%
\subsection{RGS continuum fitting}
\label{xcont}
We show the composite RGS spectrum of \th \source \th
in Fig. \ref{rgs_sp.fig}.
We fitted the spectrum in the 8-38 \th \AA
\th band and we applied a factor 5 binning. 
Because the (more appropriate)
C-statistic cannot be computed in the case of our
combined fluxed spectrum, 
for the fit we used the $\chi^2$ statistic. We
checked that at least 20 counts were collected
in each spectral bin, as required for using the
$\chi^2$ statistic.
We took into account the cosmological redshift and
the effects of Galactic neutral absorption;
we used the HOT model (collisional ionized
plasma) in SPEX for it, considering a column 
density of \th\nh=$1.26 \times 10^{20}$ \colc \th \citep{kal2005}
along the line of sight
and a temperature of 0.5 eV to mimic a neutral gas.\\
We modeled the
continuum emission with a broken power law.
The best fit parameters we obtained for it are
$\Gamma_1=2.73 \pm 0.02$ and 
$\Gamma_2=2.30 \pm 0.03$
as indexes,
with a break energy of
$\rm E_0=0.80 \pm 0.03 \, keV$.
\subsection{RGS emission lines}
\label{xem}
\begin{figure*}[thdp]
 \centering
 \includegraphics[angle=90,width=\textwidth]{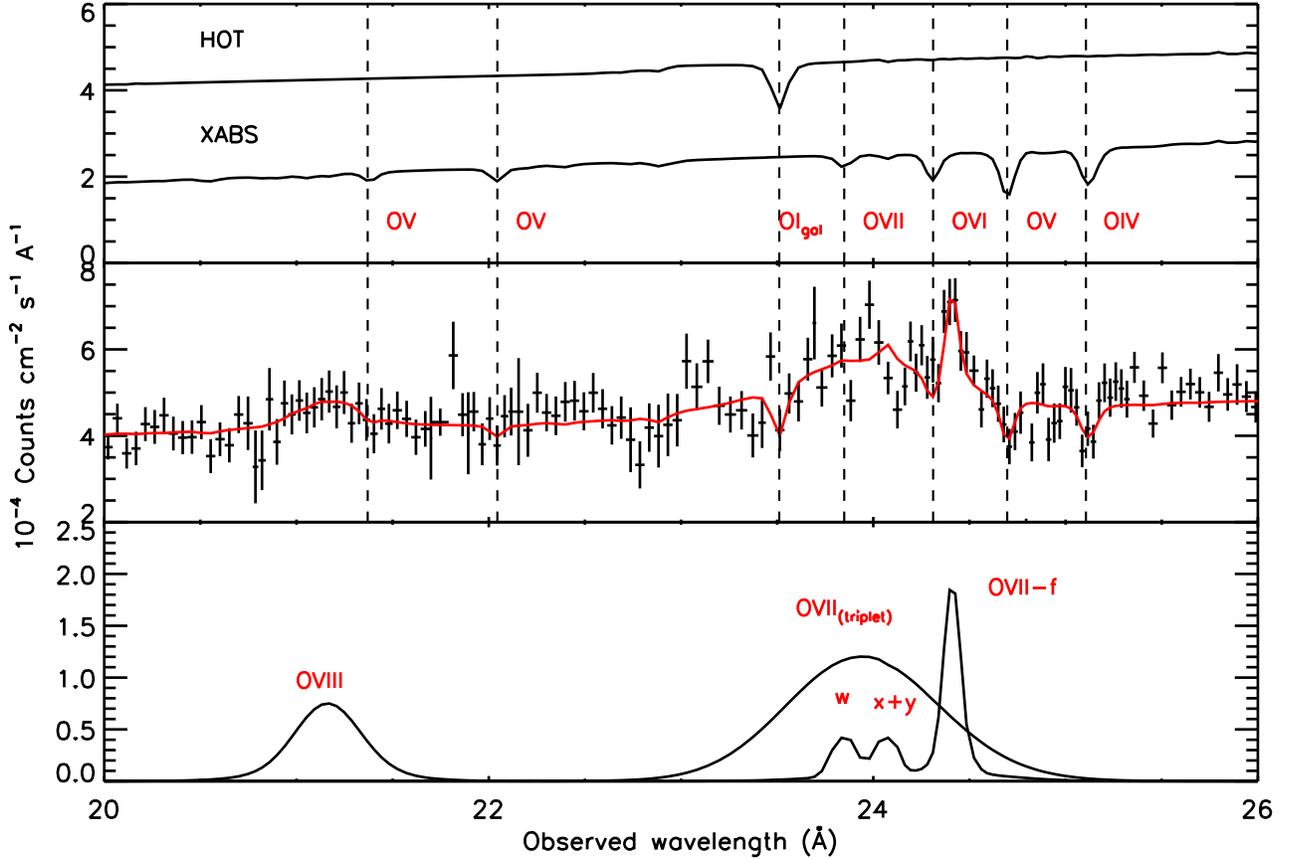}
    \caption{ \textit{Middle Panel}: best fit to the RGS spectrum 
              of \th \source \th in the 20-26 \AA\ wavelength range.
              \textit{Upper Panel}: absorbing components. The neutral
               absorbers at z=0 and the local warm absorber
               (shifted downward for plotting purpose)
               are displayed as solid lines.
               \textit{Lower Panel}: profiles of the broad and
               narrow emission features
              }
       \label{rgs_o7.fig}
 \end{figure*}

Superposed on the underlying continuum
the spectrum exhibits emission lines,
most of which are broadened. To model the broad
emission features of the spectrum, we added to the fit
a Gaussian for each candidate emission line, 
leaving the centroid, the line-width and the
flux as free parameters. We also tested the
significance of the improvement of the 
\th \chirid \th given by the extra component
through an F-test. The F-test is trustworthy
in testing the presence of an emission line
if the line normalization is allowed
to vary also in the negative range
 \citep{pro2002}.\\
We show the results in Table \ref{bl.tab}.
Three broad components gave a highly significant
improvement of the
fit. Beside the \ion{Ne}{ix} blended triplet,
and the \ion{O}{viii} \th \lia \th
line, respectively redshifted at
$\approx$ 15.1 \AA\ and
$\approx$  21.2 \AA\
(Fig. \ref{rgs_sp.fig}), 
the most prominent broad
emission feature we were able to detect
in the spectrum is a blend of the lines
of the \ion{O}{vii} triplets. In Fig.
\ref{rgs_o7.fig} we provide a zoom
on the 20-26 \AA\ spectral region
where \ion{O}{vii} and \ion{O}{viii} 
lines are seen.
To fit the spectrum
in this complex region,
we first modeled the
prominent  narrow
line visible at $\approx$ 
24.4 \AA\ with a delta
function, finding a rest wavelength
of $22.11\pm{0.08}$ \AA\ for it.
Therefore, 
we could identify this line as 
due to the forbidden transition of
\th\ion{O}{vii}\th: the line is formally 
detected with the F-test giving a
significance above 99.99\%.
The line was unresolved, 
but we could
however estimate un upper
limit for its line-width.
Assuming $\approx$ a 3:1
for the forbidden-to-intercombination
line ratio \citep{por2000}, we
provided the intercombination
and resonance lines corresponding
to the detected \ion{O}{vii}-f line.
After adding these two narrow lines, 
the spectrum was still poorly
fitted in the the 23.5-24.5 \AA,
showing broad
prominent
residuals. Hence,
we added to the fit 
a broad-line
leaving the centroid free
to vary in the range among the
known transitions 
of the \ion{O}{VII} triplets.
The modeling of the oxygen 
emission features allows a proper
detection of any intervening absorption system
\citep[e.g][]{cos2007}. Indeed,
in the same wavelength region covered
by the blend of the \ion{O}{vii} lines, transitions 
from \ion{O}{iv}-\ion{O}{vii}
are in principle present.\\
Beside the \ion{O}{vii}-f line,
we determined
upper limits for the luminosity
of several other narrow lines.
Since
the narrow line width is not resolved
in the RGS spectrum, we modeled them
with a delta function.
We estimated the upper limits
by adding to the fit a delta line 
at the wavelength of the known
transition, and fitting the maximum
normalization where the line is undetectable
over the continuum. The parameters
of the X-ray narrow lines are listed
in the upper panel of Table \ref{nl.tab}.

%
%
\begin{table}[th]
\caption{Parameters of \source \th X-ray broad lines.}
\centering
\label{bl.tab}
\begin{tabular}{lcccc}
\hline\hline 
Ion \tablefootmark{a}& Rest wavelength\tablefootmark{b} & $\log(L_{\rm obs})$\tablefootmark{c} & \fwhm  \tablefootmark{d} & F-test \tablefootmark{e}\\
& \AA & \ergs & \kms &\\
\hline
\ion{O}{vii} & $21.69\pm{0.07}$ & $ 42.44 \pm 0.07$ & $11000\pm{1000}$   & $ > 99.99\%$\\ 
\ion{O}{viii} & $19.17\pm{0.06}$ & $ 41.95 \pm 0.16$ & $5000\pm{2000}$ &  99.2\%\\
\ion{Ne}{ix} & $13.66\pm{0.05}$ & $ 41.81 \pm 0.15$ & $5000\pm{3000}$  &  99.6\%\\ 
\hline						    			      
\end{tabular}
\tablefoot{	
\tablefoottext{a}{Line transition. Note that \ion{O}{vii} and \ion{Ne}{ix} are blends.}
\tablefoottext{b}{Wavelength of line centroid in the local frame.}
\tablefoottext{c}{Intrinsic line luminosity.}
\tablefoottext{d}{Line Doppler broadening as given by the full width at half maximum of the fitted Gaussian.}
\tablefoottext{e}{Line significance as given by the F-test probability.}	
}							    
\end{table}
%

\subsection{RGS absorption lines}
\label{abs}
We modeled the X-ray absorbing gas of \th \source \th
using the photoionized-absorption model XABS in SPEX.
This model calculates the transmission of a slab of
material, where all ionic column densities are linked
through the ionization balance. Thus, it computes 
the whole set of absorption lines produced by a photoionized 
absorbing-gas, for a gas column density \th \nh \th
and an ionization parameter $\xi$.
The gas outflow velocity $v_{\rm out}$
is another free parameter.
We kept instead the RMS velocity broadening of the gas
frozen to the default value (100 \kms).
The input ionization balance for XABS is 
sensitive to the spectral shape of ionizing SED:
we calculated it running the tool \th \xabsinput, 
with the SED described in Sect. \ref{sed}
as input.
The \th \xabsinput \th routine, 
makes internally use of the Cloudy code
\citep[ver.10.00,][]{fer1998}
to determine the ionization
balance. In the calculation
we assume solar abundances
as given in Cloudy 
(see Cloudy manual 
\footnote{\url{http://nublado.org/}}
for details).
We found that the absorption features of \th \source \th
can be fitted by a thin and lowly-ionized 
absorber. The best fit parameters are
\th\nh=$8 \pm 3 \times 10^{19}$ \colc and
$\log\xi=0.03 \pm 0.15 $.
 We had no
strong constraints on a possible outflow
velocity. We estimated an upper limit
at 99\% confidence
($\Delta \chi^2$=6.67 for one parameter)
of $v_{\rm out} \leq 210$ \kms. By adding
the absorption component we achieved
an improvement of the statistic of $\Delta \chi^2$=25.
According to the F-test,
given the 3 extra degrees of freedom,
this $\chi^2$ improvement is significant at a 99.7\% 
level of confidence.
The most prominent
absorption lines (Fig. \ref{rgs_o7.fig}) are transitions
from lowly ionized oxygen species such as \ion{O}{iv}
and \ion{O}{v}, respectively redshifted to
25.1 \th \AA\ and
24.7 \th \AA\ (Fig. \ref{rgs_o7.fig}).
In Table \ref{abs.tab} we provide the column densities of
the main ions of the absorber, provided by XABS. 
As a comparison, we fitted also each ionic column density 
individually with the SLAB model. SLAB is a simpler 
absorption model where all the ionic column densities
are modeled independently, since they are not linked each 
other by any photoionization model. In the line-by-line 
fit with SLAB, we kept the same outflow velocity  and velocity
broadening of the previous XABS fit.
XABS predictions
and SLAB fit are consistent
within the quoted errors. Therefore, hereafter
we will use the value provided by XABS as reference.
\begin{table}[htdp]
\caption{Column densities of the main absorbing ions for \th \source \th.}
\centering
\begin{tabular}{lcc}
\hline\hline 
& XABS \tablefootmark{a}& SLAB \tablefootmark{b} \\
Ion & $ \rm \log(N_{ion})$ & $\rm \log(N_{ion})$ \\
& \colc & \colc \\
\hline
\ion{O}{iv}   & $16.2 \pm{0.1} $ & $ 16.1 \pm 0.2 $ \\	  
\ion{O}{v}    & $16.2 \pm{0.1} $ & $ 16.4 \pm 0.4 $\\	  
\ion{O}{vi}   & $15.9 \pm{0.1} $ & $ \leq\,15.6 $\\	  	  
\ion{O}{vii}  & $15.2 \pm{0.1} $ & $ 15.9 \pm 0.5 $ \\	  
\ion{C}{v}    & $16.1 \pm{0.1} $ & $ \leq \,16.8 $ \\	   	  
\ion{N}{iv}   & $  15.3 \pm{0.1} $ & $ \leq 15 $ \\	           
\ion{N}{v}    & $  15.3 \pm{0.1} $ & $ 15.4_{-1.7}^{+ 0.4}$ \\	   
\ion{N}{vi}   & $  15.3 \pm{0.1} $ & $ \leq 15.5$  \\	  	   
\hline
\end{tabular}
\tablefoot{
\tablefoottext{a}{Ionic column densities provided by the XABS model. 
Quoted errors are from the propagation of the error on the fitted hydrogen column density.}
\tablefoottext{b}{Ionic column densities from the line by line fitting performed with the SLAB model.}
}
\label{abs.tab}
\end{table}
%
%
\subsection{The UV and optical narrow emission lines}
\label{optem}
 \begin{figure}[thdp]
 \centering
 \includegraphics[angle=-90,width=0.5\textwidth]{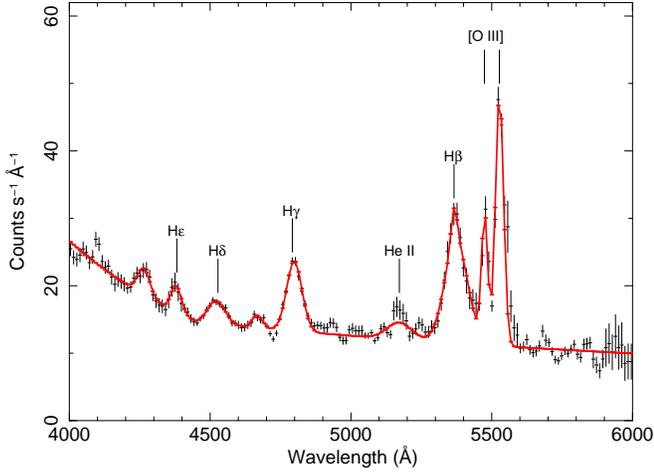}
    \caption{OM optical grism spectrum of \th \source \th in the 4000-6000 \AA\ range.
             The best fit model is displayed by a solid line.
             The main emission features 
             from the hydrogen Balmer series an the [\ion{O}{iii}] doublet are labeled.}
       \label{omgrism.fig}
 \end{figure}


\begin{table}[t]
\caption{Parameters of  the X-ray, UV, and optical  narrow lines of \th \source \th.}
\label{nl.tab}
\centering
\begin{tabular}{lccc}
\hline\hline 
Ion & Wavelength \tablefootmark{a} & $\log(L_{\rm obs})$ \tablefootmark{b} & \fwhm  \tablefootmark{c}\\
& \AA & \ergs & \kms\\
\hline  
\ion{C}{vi}   & 33.73  &  $\leq 41.67$ & ... \\ 
\ion{N}{vi}   & 29.53  &  $\leq 40.73$ & ... \\
\ion{N}{vii}  & 24.78  &  $\leq 40.65$ & ...  \\
\ion{O}{vii}-f & $22.11\pm{0.08}$ & $ 41.84 \pm 0.08$ & $\leq 2500$\\  
\ion{O}{viii}   & 18.97  &  $\leq 41.69$ & ... \\ 
\ion{Ne}{ix}    & 13.70  &  $\leq 40.52$ & ...  \\ 
\hline
\ion{C}{iv} & 1551 & 42.89 & $805 \pm 40$\\  
\ion{C}{iv} & 1548 & 43.03 & $805 \pm 40$\\  
\lia & 1215 & 43.18 & $488 \pm 24$ \\  
\ion{O}{vi} & 1038 & 42.69 & $840 \pm 42$\\  
\ion{O}{vi} & 1032 & 42.73 & $840 \pm 42$ \\ 
\hline
$[\ion{O}{iii}]$ & 5007 & $43.13\pm0.01$ & $1200 \pm 100$ \\
$[\ion{O}{iii}]$ & 4959 & $42.71\pm0.03$& $500 \pm 300$  \\
H${\beta}$ & 4861  & $42.14\pm0.12$ & 488 (fixed)  \\
\hline					    			      
\end{tabular}									     
\tablefoot{
Parameters and errors for the X-ray, UV and optical narrow emission-lines of \th \source \th.
\tablefoottext{a}{Wavelength of the transition in the rest frame.} 
\tablefoottext{b}{Intrinsic line luminosity.} 
\tablefoottext{c}{Doppler broadening of the line, as given by the full width at half maximum of the fitted Gaussian. 
Note that the X-ray lines are not resolved in the RGS spectrum.}
}
\end{table}

We studied also the narrow emission-lines 
present both in the optical
and in the UV spectrum. 
In the middle panel of Table \ref{nl.tab},
we list the UV narrow emission -lines derived from
the fit performed on the simultaneous HST-COS spectrum.
Each line was fitted by a combination 
of resolved broad and narrow components
(see Sect. 3.1 and Fig. 2 
in E11 for details).
From the parameters of the fit,
a formal 5\% error was estimated on the
narrow component,
both for the line luminosity
and for the line width.
This error does not include
any uncertainty associated
to the blend of the narrow and
the broad components.\\
Finally, in the lower panel of Table \ref{nl.tab}
we list the optical narrow emission-lines,
obtained 
by fitting the OM grism spectrum (Fig. \ref{omgrism.fig}).
We fitted
the continuum with a broken power-law,
and we modeled the sinusoidal
feature due to a residual Modulo-8 fixed pattern noise
with Gaussians. 
We clearly
detected the broadened features
of the hydrogen Balmer series
(H${\beta} \,\rm \lambda4861$ \AA, 
H${\gamma}\,\rm \lambda4341$ \AA,
H${\delta}\,\rm \lambda4102$ \AA\ and
H${\epsilon}\,\rm \lambda3970$ \AA)
and the narrow emission
lines of the [\ion{O}{iii}] doublets
$\rm \lambda \lambda$5007,4959 \, \AA. 
In the analysis we used the luminosities of
the [\ion{O}{iii}] lines and of the narrow component
of H${\beta}$. 
We modeled the [\ion{O}{iii}]
lines with Gaussian profiles, 
leaving the
centroid, the line-width
and the flux as free
free parameter.
The full width at half maximum
(\fwhm) of the 
[\ion{O}{iii}] lines was resolved
and it is reported in Table \ref{nl.tab}.
In fitting the H${\beta}$ line,
we used two Gaussians, respectively 
for a narrow and a broad component. We set the
wavelength of the narrow component to the nominal
wavelength of the H${\beta}$ transition and its
width to the one of the \th \lia \th narrow component
measured in the UV, leaving only the
normalization as free paramaters.
All the parameters of the broad component
were instead left free. The fitted
\th \fwhm \th for the broad H$\beta$ component
($4700\pm400$ \kms) is within the range
of the three broad components of
\th \lia \th measured in the UV spectrum
($\approx$ 1000, 3400 and 14\,000\th \kms).
%
%
\section{Photoionization modeling}
\label{cfr_em}

 \begin{figure}[t]
 \centering
 \includegraphics[angle=90,width=0.5\textwidth]{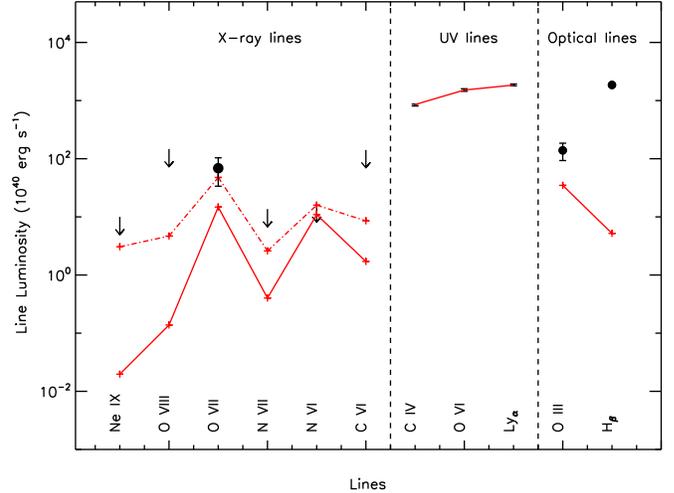}
    \caption{Best fit of the intrinsic luminosities of the UV narrow emission lines (\textit{middle panel}).
             The model is displayed by a solid line only to guide the eye.
             The line luminosities predicted by the model in the X-ray (\textit{left panel}) 
             and in the optical (\textit{right panel}) band are displayed by red crosses connected by a solid line.
	     The model agrees with the upper limits for the X-ray lines, but does not fit the optical lines.
	     The qualitative best fit model for the X-ray lines is also shown by red crosses connected by a dot dashed line.
             Error bars, when larger than the size of the plotting symbols, are also shown.
             Note that a 3$\sigma$ error bar is shown for \ion{O}{VII}-f line.}
       \label{fitcloudy.fig}
 \end{figure}

%

\begin{table}[t]
\caption{Covering factor, column density and ionization parameter for 
the gas emitting the UV and X-ray narrow lines in \th \source \th.}
\centering
\begin{tabular}{lccc}
\hline\hline
 & $C_v$ & $\rm \log \nh$ & $\rm \log \xi$ \\
 & &\colc & \\
\hline 
UV \tablefootmark{a}& $0.104 \pm 0.006 $ & $21.67 \pm 0.25$  & $0.48 \pm 0.15$ \\
X  \tablefootmark{b}& 0.104 (fixed) & $21.40 \pm 0.25$  & $1.44 \pm 0.15$ \\
\hline
\end{tabular}
\tablefoot{
\tablefoottext{a}{Best fit parameters for the gas emitting the UV lines.}
\tablefoottext{b}{Parameter of the gas model consistent with the observed upper limits and
best fitting the only detected X-ray line. We imposed the same gas covering factor of the UV-emitter.
See text for details.}
}
\label{fitcloudy.tab}
\end{table}
%

To estimate the global properties of the gas emitting
the narrow lines in \th \source,
we used the photoionization code Cloudy v. 10.0 \citep{fer1998}.
In the calculation
we assumed total coverage of the source
and a gas density of $\rm 10^8 \, cm^{-3}$.
We note that the assumption
on the gas density is not critical
for the resulting gas parameters. Indeed
line ratios of He-like ions are not sensitive to
the gas density over a wide
range of density values \citep{por2000}.
We used the SED described in Sect. \ref{sed}
and the ionizing luminosity derived from it
as input. 
We created a grid of line luminosities for
a wide range of possible gas parameters:
the column density ranged between $\rm \log \nh = [19-23] \, cm^{-2}$,
and the ionization parameter between
$\rm \log \xi = [-0.3-2.7] \, erg \, cm \, s^{-1}$, 
with a spacing of 0.25 dex and 
0.15 dex respectively.
In the fitting routine,
we first computed for each grid entry
the gas covering factor $C_v$, 
by imposing to the model
match the luminosity of the \th \lia \th line.
Hence, we modeled the data by minimizing
the merit function:
$$
\chi^2=\sum \frac{(C_v L_c-L_{\rm obs})^2}{\sigma_{\rm obs}^{2}},
$$
where
$L_c$ is the model-predicted luminosity of each line,  
$C_v$ is the gas covering factor 
and $L_{\rm obs}$ is the observed line luminosity, with statistical error $\sigma_{\rm obs}$.
We fitted all the UV, X-ray and optical luminosities listed
in Table \ref{nl.tab}.
The \ion{C}{iv},\th \ion{O}{vi} and [\ion{O}{iii}]
doublets were considered as a blend and
$3\sigma$ upper limits for the X-ray 
lines of Table \ref{nl.tab} were also included.\\
A model with a $\approx$ 10\% covering factor 
$\rm \log \nh \approx 21.67$ \colc and 
$\log \xi \approx 0.48$
fits the three UV data points
(\lia, \ion{C}{iv}, \ion{O}{vi})
and agrees with all the upper
limits for the X-ray luminosities.
In fitting the UV lines 
we obtain a minimum $\chi^2/\rm{d.o.f}=0.35/1$.
With just one degree of freedom 
the probability of getting such a low
observed $\chi^2$ or lower for a correct model 
is 0.45.The derived
parameter for the UV emitter
are outlined in Table \ref{fitcloudy.tab}.
We report the width of the grid step as
error on the parameters. The error
for the covering factor is estimated
propagating the error ($\approx$5\%) 
on the luminosity
of the \th \lia \th line.\\
This best fit model 
neither match the luminosity of the \ion{O}{vii}-f
nor the luminosities of the optical lines,
suggesting that X-ray and optical
lines arise in a gas with different
physical conditions.
It was not possible to
estimate the parameters
of the X-ray emitter
through a proper fit,
because we had only
one line detection. However,
the measured upper
limits provide useful
constraints for the model.
Assuming the same covering factor
of the UV emitter, we selected 
in the grid all the models
in agreement with the measured upper
limits and matching the luminosity 
of the \ion{O}{vii}-f line within
3$\sigma_{\rm obs}$. In this calculation, 
we added
the contribution of the
best-fit UV model to the
X-ray luminosities.
We find that the selected models
have
$\log \nh$ ($[21.13-21.93]$ \colc) and
their ionization
parameter is comprised
within 
$\log \xi=[1.12-2.06]$.
Therefore, the fit which best
model the \ion{O}{vii}-f 
line has a column
density consistent, within
a grid step, with the UV emitter one
($\log \nh \approx$21.40 \colc) but
has a higher ionization parameter
($\log \xi \approx$1.44). This
may indicate a geometrical
connection between the UV and the
X-ray emitter. In the case
when also the gas covering factor
is left free, we obtained a larger
parameter range, 
($C_v=[0.03-0.60]$, 
$\log \nh=[20.90-23.00]$ \colc), still
indicating a
higher ionization
parameter 
($\log \xi=[1.12-1.60] $) with respect to the
UV-emitter.
We remark also that none
of the models in our grid
match the optical-lines luminosities.
Finally we note that models
with parameters consistent
with the warm 
absorber detected in the
RGS spectrum neither fit
the UV ($\chi^2=[246-631]$) nor the
the X-ray (the \ion{O}{vii}-f
is not fitted within $\approx 6\sigma$) 
emission lines. We show the
best fit models for the UV and the X-ray emission 
lines in Fig. \ref{fitcloudy.fig}. The derived
parameter for the X-ray emitter
are outlined in the second
row of Table \ref{fitcloudy.tab}.
%
%

%
\section{Comparison between the UV and the X-ray absorber}
\label{cfr_abs}
We investigate the possible relationship
between the UV absorber found for
\th \source \th in E11 and the X-ray absorber
found here (Sect. \ref{abs}) by comparing
the gas parameters independently measured
in the two different wavebands.\\
Three different outflow components 
($\rm v_1=-38 \,  \kms$, 
$\rm  v_2=-156 \,\kms$
and $\rm  v_3=-220 \, \kms$) 
were identified in the UV.
Because of the
the lower RGS resolution,
we were not able to resolve any of the UV
components; however our upper
limit for the outflow velocity of the X-ray absorber 
is consistent with all of them.
Thus, it is likely that in the X-rays we observed a blended 
superposition of the UV components.\\
In Table \ref{cfr_abs.tab} we compared
of the UV and  X-ray column densities for
\ion{C}{ii}, \th \ion{C}{iv}, \th \ion{N}{v}, \th \ion{O}{vi}. Absorption
lines from \ion{O}{vi} were detected in both bands,
while lines from \ion{C}{ii}, \ion{C}{iv} and \ion{N}{v} were
detected only in the UV; however
their column densities could be
predicted by our XABS model. 
We report in Table \ref{cfr_abs.tab} 
the sum of the column densities of
the kinetic components 1, 2 and 3,
obtained using a UV partial
covering model, with a power
law distribution of the
optical depth (see E11 for details). 
The column densities found independently 
in the X-ray and in the UV
agree within the quoted errors for all ions.\\
The ionization parameter of the UV absorber is
given in terms of $ {U_H=\frac{Q_H}{4 \pi R^2 c \nh}}$ 
( where $Q_H$ is the rate of hydrogen ionizing photons emitted by the source,
$R$ is the absorber distance from the source,
c is the speed of light,
and \th \nh \th is the total hydrogen column density)
rather than in terms of $\xi$.
For the SED of \th \source, 
$\log \xi=1.7+ \log U_H$.
In the UV the determination of $U_H$ is not
well constrained:
for a broken power-law SED
and for a gas with solar abundances
$\log U_H$ =[-1.7 to -1.5]
(Fig. 8 in E11). Applying the conversion
factor just given this  value corresponds to
$\rm \log \xi$ =[0 to 0.2], consistent
with the more constrained value 
found here for the X-ray
absorber: $\rm \log \xi$ =[-0.12 to 0.18].
%
\begin{table}[htdp]
\caption{Comparison between the UV and X-ray column densities of the absorbing ions in \th \source \th.}
\centering
\begin{tabular}{lcc}
\hline\hline
& UV \tablefootmark{a} & X \tablefootmark{b} \\ 
Ion & $\rm \log(N_{ion}) $ & $\rm \log( N_{ion})$ \\
& \colc & \colc \\
\hline
\ion{C}{ii}   &	 $12.99-13.08$     & 13.11 \\
\ion{C}{iv}   &  $15.16-15.29$ & 15.42  \\ 
\ion{N}{v}    &  $15.16-15.31$ & 15.36 \\ 
\ion{O}{vi}   &  $15.83-15.96$ & 15.89 \\ 
\hline 
\end{tabular}
\label{cfr_abs.tab}
\tablefoot{
\tablefoottext{a}{Ionic column densities measured in the UV.
We considered the sum of the three kinetic components detected.}
\tablefoottext{b}{Ionic column densities predicted the X-ray model (XABS), discussed in Sect.\ref{abs}.}
}
\end{table}
%

\section{Discussion}
\label{disc}

\subsection{Absorber distance and energetics}
The spectral analysis of the RGS spectrum of
1H0419-577, together
with the analysis of the simultaneous COS spectrum 
(E11) revealed that 
the X-ray and the UV emission spectra of \th \source \th
are both absorbed by a thin, weakly ionized
absorber. As pointed out in Sect. \ref{cfr_abs} 
the ionization parameter and the ionic column densities
measured independently in the UV and in the X-ray agree
with each other.
Thus, it is likely that the UV and X-ray
warm absorber are one and the same gas. 
In the UV the kinetic structure of the
warm absorber is resolved, with three
outflowing components detected.
The bulk of  absorbing column density 
is carried  by the components 2+3 (see Table 1 and 2 and
Fig. 3 in E11). Therefore
in the lower resolution X-ray spectrum we likely 
detected a blend of the UV kinetic
components dominated by the UV components 2+3.
Having established the connection between the
UV and the X-ray absorber, we can exploit
the complementary information derived
in the two wavebands to estimate the
warm absorber location
and energetics.
\\ 
In the following discussion
we consider that an ionized medium, 
parameterized by
$\rm \log \xi \approx 0.03 $ 
and \nh $\approx 8  \times 10^{19}$ \colc
(this paper), 
is outflowing from the source at the velocity
of the UV component 3 ($\rm  v_3=-220 \, \kms$).
Additionally, we use a a gas number density
$n_{\rm H}\leq\,25\,\rm cm^{-3}$, as estimated
in E11 for the UV component 2+3.
This upper limit for the gas number density
was derived from the ratio
of the (non detected) 
first excited metastable and the ground
level of \ion{C}{ii}. As already
pointed out in Sect. \ref{cfr_abs},
the \ion{C}{ii} column density measured in the
UV is well accounted for by
our absorber model.
We note however that in the
COS spectrum, the
\ion{C}{ii} 1334.5 \AA\ profile 
is possibly asymmetric, and the centroid 
is slightly shifted with
respect to the UV component 3:
at least part of the \ion{C}{II}
emission may in principle 
arise in a kinetic component
separated from the UV component 3.
However, this possible 
additional component is not evident 
in any other lines.\\
Given the X-ray UV connection just
established, the following discussion is
based on the assumption that
the upper limit for the gas density 
estimated in the UV
may be applied to the UV/X-ray absorber.
We assume a thin shell geometry for
the outflow. In this
approximation, the outflow is
spherically shaped, with a global
covering factor of the line of sight
$C_g$.
Each outflowing shell,
$\Delta r$ thick, is partially filled with gas,
with a volume filling factor $f$.
The volume filling factor $f$ can be 
estimated analytically \citep{blu2005},
from the condition that the kinetic momentum of the outflow must
be of the order of the momentum of the absorbed
radiation plus the momentum of the scattered radiation.
Applying this condition, we found
that the absorber in \th \source \th has a volume filling factor
$f \approx \rm 3 \times 10^{-3}$, suggesting that it may
consist of filaments or fragments very diluted in the
available volume and intercepting our line of sight.\\
Exploiting the tighter constraint
on the ionization parameter provided by
the present analysis we confirm
the distance estimation given in E11:
\begin{equation}
\label{rlw2}
R \geq (\frac{L_{\rm ion}}{n_{\rm H} \xi})^{1/2} \ga 3 \, \rm kpc.
\end{equation}
This estimation places the warm absorber
at the host galaxy scale,
well outside the central region with the
broad-line region 
\citep[$R_{\rm BLR} \approx 0.07 \rm \,pc$,][]{tur2009}.
UV absorbers located
at a galactic-scale distance are not uncommon
in low-redshift quasars (see table 6 in E11 and
references therein). 
The X-ray/UV connection
we infer for this source
would therefore make 
its low-$\xi$
absorber
the first galactic scale X-ray
absorber ever detected.
A possible confining medium
for an X-ray absorbers
located at a $\approx$kpc
from the nucleus could
be a radio jet-like
emission. This source
is radio quiet, but
a 843 MHz flux detection
\citep{mau2003} may
be due to a weak radio
lobe. More accurate
radio measurements 
are required
to test 
this possibility.
Such a galactic-scale wind may be both
AGN or starburst driven. However, for
this source, the UV analysis suggests that the
photoionization of the outflow may
be dominated by the AGN emission
(see discussion in E11
and references therein).\\
We estimated also an upper limit 
for the WA distance
from the condition that
the thickness $\Delta r$ of the outflowing-gas column 
should not overcome its distance $R$ from the centre
\citep{blu2005}.
Analytically, this condition is:
\begin{equation}
\label{rup1}
\frac{\Delta r}{R} \approx \frac{\nh}{ f n_H R} = \frac{\xi R \nh}{L_{\rm ion} f} \leq 1 .
\end{equation}
From Eq. \ref{rup1} we derived: $R \la 15 \rm \, kpc$.
We note that this distance
is well within the typical
extension of a galactic halo
\citep[e.g as mapped by H I emission 
for a large sample,][]{deb2008}.
Therefore we use this estimation
in the following to derive
hard upper limits for the mass outflow rate 
and kinetic luminosity.
\\
The outflow mass rate
is given by
\begin{equation}
\label{mtot}
 \dot{M}_{\rm out}=4 \pi \mu m_P R v \nh C_{g}=[4-16] \, \msunyr,
\end{equation}
where $\mu$=1.4 is the mean atomic mass per proton,
$m_P$ is the proton mass. We assumed a covering
factor $C_g$=0.5, given by the fact that outflows
are seen in about 50\% of the observed Seyfert
galaxies \citep{dun2007}.
\\
This value may be compared with the classical
mass accretion rate of a black hole in the Eddington regime
($\dot{M}_{\rm Edd}=L_{\rm bol}/\eta c^2$) to obtain
an estimation of the impact of the mass loss
due to the outflow on the AGN. Assuming
a typical accretion efficiency $\eta=0.1$ and 
taking $L_{\rm bol}=9 \times 10^{45} \ergs$,
as estimated from the SED (see Sect. \ref{sed}) 
we obtained
\begin{equation}
\dot{M}_{\rm Edd}\approx 2 \msunyr.
\end{equation}
As found in most cases \citep[see][]{cos2010},
the mass outflow rate can be of the
same order of the mass accretion rate,
suggesting a balance between accretion and 
ejection
in this system.
The kinetic luminosity of the outflow is: 
\begin{equation}
L_{\rm kin}=\frac{\dot{M}_{\rm out} v^2}{2} \approx 10^{40.7-41.4} \, \ergs,
\end{equation}
and it represents a 
small fraction
($\la 10^{-2}$\% )
of the AGN bolometric
luminosity $L_{\rm bol}$; thus the outflow is
not energetically significant in the AGN feedback 
scenario, where kinetic luminosities
of a few percent
of the bolometric luminosities
are required \citep{sca2004}.
We finally estimated the maximum kinetic energy that
the outflow can release into the 
interstellar medium, in the case it
is steady all over the AGN life time
\citep[$\approx 4 \times 10^8 \rm \,yr$,][] {ebr2009}:
\begin{equation}
E_{\rm tot} \approx 10^{56.8-57.5}\,\rm erg.
\end{equation}
As argued in \citet{kro2010} this value
may in principle be sufficient
to evaporate the interstellar
environment out of the host galaxy.
However, it is not trivial to couple
this energy effectively to the
galaxy \citep[e.g.][]{kin2010}.
\subsection{The origin of the emission lines}

The simultaneous \th \Cos \th
and \th \xmm \th observation
of \th \source \th provided
a set of narrow lines, 
ranging from the optical to the
X-ray domain, suitable for photoionization
modeling. We show that a single gas model
cannot account simultaneously 
for all the narrow-lines emission. 
The UV lines are emitted
by moderately ionized
gas, intercepting about the 10\%
of the total AGN radiation field.
This value for the covering factor
is consistent with what previously
reported
\citep[$C_v$=1.9\%--20.5\%,][]{bas2005}.
The X-ray lines are instead 
emitted in a more highly 
ionized gas phase: ($\log \xi \approx 1.44$). 
We also found that
a gas with the same 
column density and covering factor
as the UV emitter is a good description
of the X-ray emission.
This may suggest that
the two emitters are two
adjacent
layers of the same gas.\\
Most of the optical emission
is not accounted for by our model:
it can explain only up to
the 4\% of the H${\beta}$ 
luminosity and the 0.3\% of
[\ion{O}{iii}] luminosity.
Lower densities are
required to emit the [\ion{O}{iii}] lines:
the [\ion{O}{iii}] $\lambda5007$\AA\ line
is indeed collisionally de-excited for
$n_H \ga 10^5 \, \rm cm^{-3}$ \citep{ost1989}.
Our simple photoionization model
cannot however account for the variety
of gas physical conditions occurring
in the narrow-line region.
Our analysis suggests that
the NLR is a stratified environment
hosting a range of different 
gas components. Previous studies have shown that
multi-component photoionization
models are required for describing the narrow
emission lines spectrum
of AGN. The narrow line
emission from the infrared
to the UV is well
reproduced assuming that the
emitting region consists 
of clouds with a wide range
of gas densities and ionization
parameters \citep{fer1997}.
In the case of NGC 4151
more than one gas component,
with different covering factor,
are required to explain
lines-emission even limiting
the analysis to the soft
X-ray regime \citep{arm2007}. In the
present case, the data
quality did not allow us
to test a more complex,
multi-component
scenario.\\
The estimated gas parameters of the warm 
absorber are 
largely inconsistent with
the emitters.
Thus, neither the UV nor the X-ray
emitter can be regarded
as the emission counter-part
of the warm absorber.
Therefore, 
a connection between the
warm absorber and the gas
in the NLR is discarded
in the present case.
\subsection{The geometry of the gas}

The present analysis of the UV and X-ray
spectrum of \th \source \th revealed
three distinct gas phases: the UV/X-ray
warm absorber, the X-ray emitter 
($\log \xi_{\rm X} \approx 1.44 $)
and the UV emitter ($\log \xi_{\rm UV} \approx 0.48$).
We used the line width of Table \ref{nl.tab}
to estimate qualitatively the location of the emitters.
Assuming that the NLR gas is moving in random 
keplerian orbits with  an isotropic velocity
distribution, the velocity $v_{FWHM}$ by which
the narrow lines are broadened
is given by \citep{net1990}:
\begin{equation}
v_{FWHM}=\sqrt{\frac{4GM}{3R}},
\end{equation}                                                                                                                     
where $M$ is the mass of the SMBH
and $R$ is the radial distance
from it. The \th \fwhm \th
of the UV lines
([488--805] \th \kms \th, Table \ref{nl.tab})
give therefore the 
approximate location of the UV emitter:
\begin{equation}
R^{\rm em}_{\rm UV} \approx [1-3] \rm \,pc.
\end{equation}
This distance would imply a gas number 
density of 
$n_{H}=(L_{\rm ion}\xi_{UV}) / (R^{\rm em}_{\rm UV})^2 \approx 10^{7-8}\, \rm cm^{-3}$
at the location of the UV emitter,
consistent
with the range of gas density and distance
where these lines are optimally emitted
\citep{fer1997}.
The X-ray emitter is consistent
to be a gas with covering factor and
column density similar
to the UV emitter. This suggests
that it may be a layer of gas
adjacent to UV emitter.
The higher X-ray ionization
parameter may
by produced both by a smaller distance 
to the SMBH
and by a lower gas density.
The gas density in
the NLR ionization cone
has a smoothly decreasing
radial profile \citep{bia2006}.
Therefore, if the UV and X-ray emitter
are not radially
detached, as we suggested,
a similar gas density
for both the emitters can be
assumed.
From the definition
of ionization parameter, it follows that:
$$
R^{\rm em}_{\rm X} \approx \sqrt{ \frac{\xi_{\rm UV}}{\xi_{\rm X}}} R^{\rm em}_{\rm UV} \approx [0.3-1.0] \, \rm pc.
$$
This estimation is consistent with the lower
limit for the distance ($R^{\rm em}_{\rm X} \ga 0.1 \rm \,pc$) 
implied by the upper limit for the
broadening of the \ion{O}{vii}-f line.
This first order estimation places
the emitters on a very different distance scale
compared to the absorber, again arguing against
a connection between emission and absorption
in this source. We did not detect
the NLR in absorption: this may indicate
that the NLR ionization cone is not
along our line of sight,
as this would produce visible
\ion{O}{vii} absorption lines.
We are possibly
detecting scattered light
from the NLR. The presence
of a circumnuclear scattering region
has been proposed, for example, for the
case of NGC 4151 \citep{kra2000}.
%
\section{Summary and conclusions}
\label{sum}
We analyzed and modeled the X-ray, UV and optical data
of the Seyfert 1.5 galaxy \th \source. 
Simultaneous X-ray (RGS) and UV (HST-COS) 
spectra
of the source were taken, to study the
absorbing-emitting photoionized gas in this source.
Optical data from the OM
were also used for the present
analysis.
We found three
distinct gas phases with
different ionization.\\ \\
The X-ray and the UV spectrum are both
absorbed by the same, lowly ionized
warm absorber ($\log\xi=0.03 \pm 0.15$). 
The outflow is likely to be located
in the host galaxy, at a distance
$R \ga 4 \, \rm kpc$
from the central source.
The kinetic luminosity of the outflow
is small fraction ($ \la 10^{-2}$\%) 
of the AGN bolometric luminosity, making 
the outflow unimportant for the
AGN feedback. However, such a galactic-scale
X-ray absorber, like the one
we serendipitously discovered in this source, 
might still play a role in the host galaxy evolution.
\\  \\
We performed
photoionization modeling of the
narrow lines emitter using the
available UV, X-ray and optical
narrow emission lines. The analysis indicates that
the narrow-lines emitters
are not the emission counter-part of the WA.
A connection between
the WA and the NLR can therefore be discarded
in this case. \\ \\
The X-ray emission
lines are emitted in a more highly ionized
gas phase compared to the one producing the UV lines.
We suggest a geometrical connection
between the UV and the X-ray emitter, where
the emission takes place in a single gas layer, 
located at $\approx$ pc scale distance from the
center. In this scenario, the X-ray lines
are emitted in a portion of the
layer located closer to the SMBH.\\ \\
Finally, our analysis suggests that
the NLR is a stratified environment,
hosting a range of gas phases with
different ionization and density.
%

\begin{acknowledgements}
This work is based on observations with XMM-Newton, 
an ESA science mission with instruments and contributions 
directly funded by ESA Member States and the USA (NASA). 
SRON is supported financially by NWO,
the Netherlands Organization for Scientific Research.
NA acknowledge support from NASA grants NNX09AT29G and HST-GO-11686.

\end{acknowledgements}



\end{document}